\documentclass[aps,pre,reprint,groupedaddress,amsmath,amssymb,amsfonts,floatfix]{revtex4-2}

\usepackage{float}
\usepackage{graphicx}
\usepackage[caption=false]{subfig}
\usepackage[colorlinks,linkcolor=blue,anchorcolor=blue,citecolor=blue,urlcolor=blue]{hyperref}
\usepackage{xurl}      % 更推荐，专门优化URL分行

\hypersetup{
    breaklinks=true,
}

\begin{document}

% Use the \preprint command to place your local institutional report
% number in the upper righthand corner of the title page in preprint mode.
% Multiple \preprint commands are allowed.
% Use the 'preprintnumbers' class option to override journal defaults
% to display numbers if necessary
%\preprint{}

%Title of paper
\title{Three-state coevolutionary game dynamics with environmental feedback}

% repeat the \author .. \affiliation  etc. as needed
% \email, \thanks, \homepage, \altaffiliation all apply to the current
% author. Explanatory text should go in the []'s, actual e-mail
% address or url should go in the {}'s for \email and \homepage.
% Please use the appropriate macro foreach each type of information

% \affiliation command applies to all authors since the last
% \affiliation command. The \affiliation command should follow the
% other information
% \affiliation can be followed by \email, \homepage, \thanks as well.
\author{Yi-Duo Chen}
%\email[]{Your e-mail address}
%\homepage[]{Your web page}
%\thanks{}
%\altaffiliation{}

\author{Zhi-Xi Wu}

\author{Jian-Yue Guan}
\email[Corresponding author: ]{guanjy@lzu.edu.cn}

\affiliation{Lanzhou Center for Theoretical Physics, Key Laboratory of Theoretical Physics of Gansu Province, Key Laboratory of Quantum Theory and Applications of MoE, Gansu Provincial Research Center for Basic Disciplines of Quantum Physics, Lanzhou University, Lanzhou 730000, China}%

%Collaboration name if desired (requires use of superscriptaddress
%option in \documentclass). \noaffiliation is required (may also be
%used with the \author command).
%\collaboration can be followed by \email, \homepage, \thanks as well.
%\collaboration{}
%\noaffiliation

\date{\today}

\begin{abstract}
% insert abstract here
Environmental feedback mechanisms are ubiquitous in real-world complex systems. In this study, we incorporate a homogeneous environment into the evolutionary dynamics of a three-state system comprising cooperators, defectors, and empty nodes. Both coherence resonance and equilibrium states, resulting from the tightly clustering of cooperator agglomerates, enhance population survival and environmental quality. The resonance phenomenon arises at the transition between cooperative and defective payoff parameters in the prisoner's dilemma game. 
\end{abstract}

% insert suggested keywords - APS authors don't need to do this
%\keywords{}

%\maketitle must follow title, authors, abstract, and keywords
\maketitle

% body of paper here - Use proper section commands
% References should be done using the \cite, \ref, and \label commands
\section{\label{sec_intro}Introduction}

The emergence and evolution of cooperative behavior among systems composed of selfish individuals has attracted considerable attention in complex systems science \cite{axelrod1981evolution,nowak2006evolutionary,nowak2006five,szabo2007evolutionary,nowak1992evolutionary}. Public goods play a crucial role in population evolution \cite{kummerli2010molecular,levin2014public}. As the well-known ``tragedy of the commons'' illustrates, selfish individuals tend to consume limited public resources without restraint. Inevitably, this leads the system to an unfavorable state characterized by resource scarcity and widespread defection \cite{hardin1968tragedy}.

When considering strategic feedback on the environment and payoff matrices coupled to the environmental state, the ``oscillatory tragedy of the commons'' (o-TOC) emerges across a wide range of payoff structures \cite{weitz2016oscillating}. Through the adaptive influence of individual strategies, environmental feedback can either intensify or alleviate the social dilemma \cite{hauert2019asymmetric}. The environmental update rate and spatial heterogeneity also critically influence the emergence of TOC or o-TOC in more realistic scenarios involving diverse feedback mechanisms \cite{lin2019spatial,tilman2020evolutionary,chen2025coevolutionary}. Self-organized clustered patterns of strategies and resources, emerging from localized feedback mechanisms, underlie the dynamic equilibrium states of coevolutionary systems \cite{chen2025coevolutionary,rietkerk2021evasion}. Furthermore, game state transitions effectively enhance the competitiveness of cooperators, even though defection is the Nash equilibrium across various system structures reflecting social relation feedback or localized environmental feedback \cite{hilbe2018evolution,su2019evolutionary,kleshnina2023effect,chen2025higherorder,nash1950equilibrium}.

Most studies have not considered dynamically evolving population size, which can substantially influence population competitiveness and, in turn, lead to either extinction or stable survival during coevolutionary processes \cite{huang2015stochastic,park2019population,tao2025dynamics}. In addition to the competition between cooperators and defectors, empty nodes provide a buffering effect that helps maintain the population, similar to the well-known third strategy, the ``loner'' \cite{park2023vacancies,hauert2002volunteering,szabo2002phase}. To further investigate the impact of environmental constraints on strategy evolution, we additionally consider the environmental regulation functions of individuals. In the weak dilemma regime with sufficiently high environmental capacity, the system attains a dynamic equilibrium state characterized by a high population level and a survival-favorable environment, resulting from mutualistic cooperative clustering. Survival coherence emerges at the parameter boundary between the dynamic equilibrium state and the TOC state with a barren environment, and is characterized by the optimal response of both population fraction and environmental state to the noise intensity.

\section{\label{sec_model}Model and methods}

We perform MC simulations on square lattices with $N=L \times L$ nodes. Each node $i$ is assigned a state $s_i \in \{ C, D, E\}$, corresponding to cooperator ($C$), defector ($D$), and empty ($E$), respectively. Initially, all nodes are randomly assigned to either active individuals (equal probability of $C$ and $D$) or empty states, each with equal probability. At the start of the simulations, we set initial fraction of both cooperators and defectors $x_{C0}=x_{D0}=0.25$, so the half of the nodes are empty, and the initial environment state is $n_0=0.75$. 

In each step, we randomly select a node $i$. If $i$ is active individual, it becomes an empty node with death probability 
\begin{equation}\label{eq_death_rate}
    \eta_i = \frac{1}{1+n\cdot \exp (\omega \pi_i)},
\end{equation}
where $\omega$ is the selection strength, $n \in [0, 1]$ is the environment state, and $\pi_i$ is the average payoff from individual $i$'s interactions with four nearest neighbors. Otherwise, if $i$ is an empty node surrounded by other active individuals, it will be occupied by the offspring of a neighbor, chosen with equal probability. During reproduction, an offspring adopts the opposite strategy with probability $\mu$ (mutation probability). It is important to note that $\mu = 0.5$ implies that reproduction corresponds to a random selection of strategies, while higher mutation probabilities lack physical significance. 

Active individuals play the prisoner's dilemma game with each other according to the following payoff matrix
\begin{equation}\label{eq_payoff_matrix}
    \Pi = 
        \begin{bmatrix}
            1-r & -r \\
            1 & 0
        \end{bmatrix},
\end{equation}
where $r \geq 0$ is the cost-to-benefit ratio in the donation game, a widely used form of prisoner's dilemma game. During each interaction, a cooperator provides a benefit of $1$ to its opponent at a cost of $r$, whereas defectors have no contributions. The dilemma is more pronounced as $r$ increases. And active individuals get $\pi_E=0$ from interactions with empty nodes. 

In this study, the fractions of cooperators, defectors and empty nodes of the system are defined as $x_C = N_C/N$, $x_D = N_D/N$, and $x_E = N_E/N$, respectively, where $N_s$ denotes the number of nodes with state $s$. Thus, $x_C + x_D + x_E=1$. In order to investigate the evolutionary dynamics from population (i.e., all active individuals) level, we also defined the following quantities. The cooperative fraction $\rho_C = N_C/(N_C+N_D)$ measures the cooperation level among active individuals. And the population fraction is defined as $x_P = (N_C+N_D)/N$. 

We define $N$ steps of selection and update process as a MC step (MCS). During $1$ MCS, each node has an equal chance to update its state once. The environment state is updated once after each MCS:
\begin{equation}\label{eq_env_update}
    n(t+1) = n(t) + \epsilon(\theta x_C - x_D),
\end{equation}
where $\epsilon$ is the environment update rate relative to strategy updates, and $\theta$ is the environmental recovery rate. We force restrict $n$ to $[0, 1]$ during the update process. Without loss of generality, we set $L=400$, $\omega=10$, $\theta=1$, and $r=0.3$ unless otherwise noted.

\section{\label{sec_rd}Results and discussion}

\subsection{\label{sec_rd_cr}Survival coherence resonance}

\begin{figure}%[!htbp]

    \includegraphics[width=\linewidth]{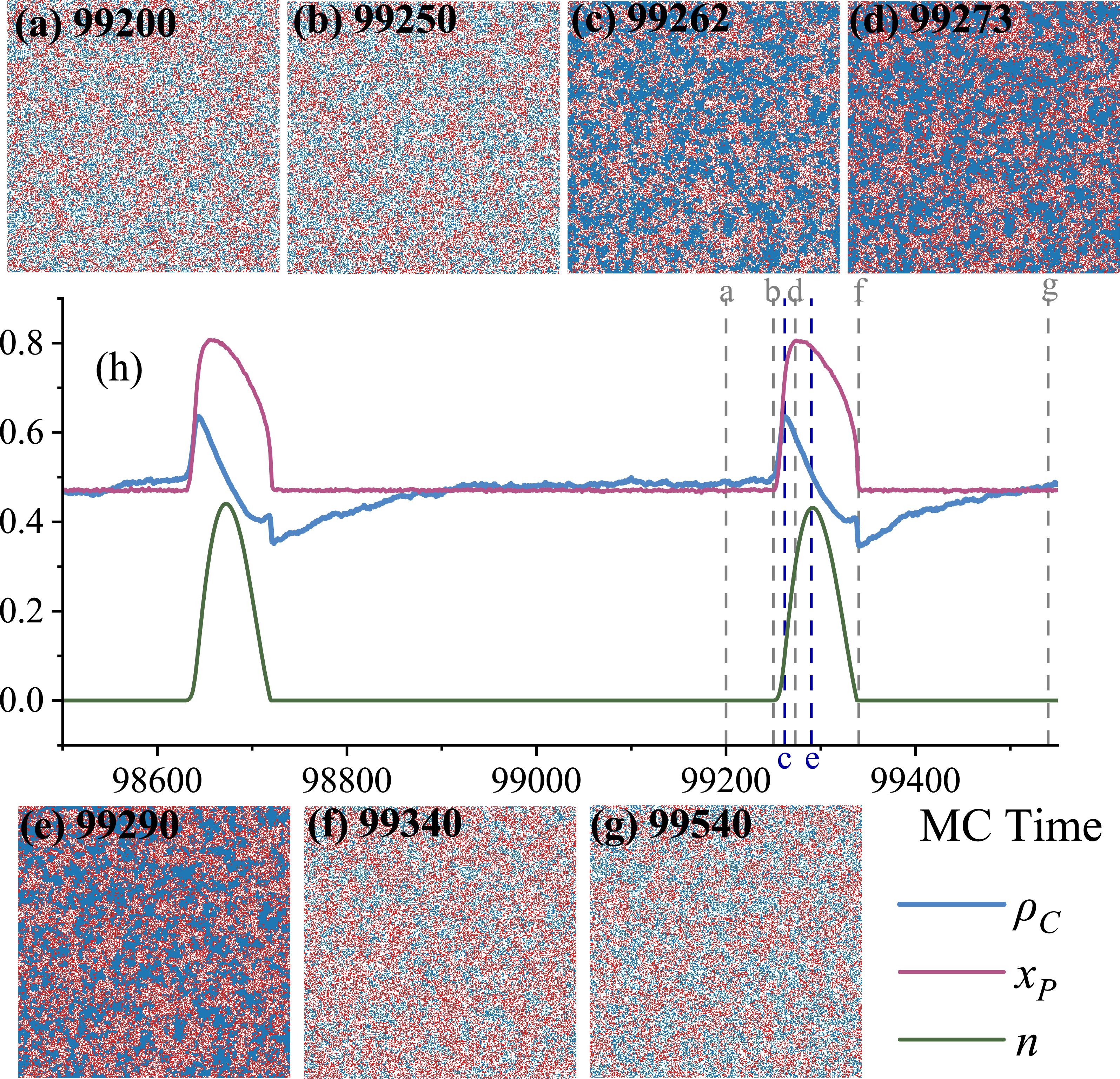}

  \caption{\label{fig_osc}Oscillatory state for $\epsilon=0.1$ and $\mu=0.005$. (a)--(g) Spatial patterns of cooperators (blue points) and defectors (red points) at different times. (h) Temporal evolution of the cooperative fraction $\rho_C$, population fraction $x_P$, and environmental state $n$. Dashed lines indicate the time points corresponding to panels (a)--(g). } 
    
\end{figure}

\begin{figure}%[!htbp]

    \includegraphics[width=0.8\linewidth]{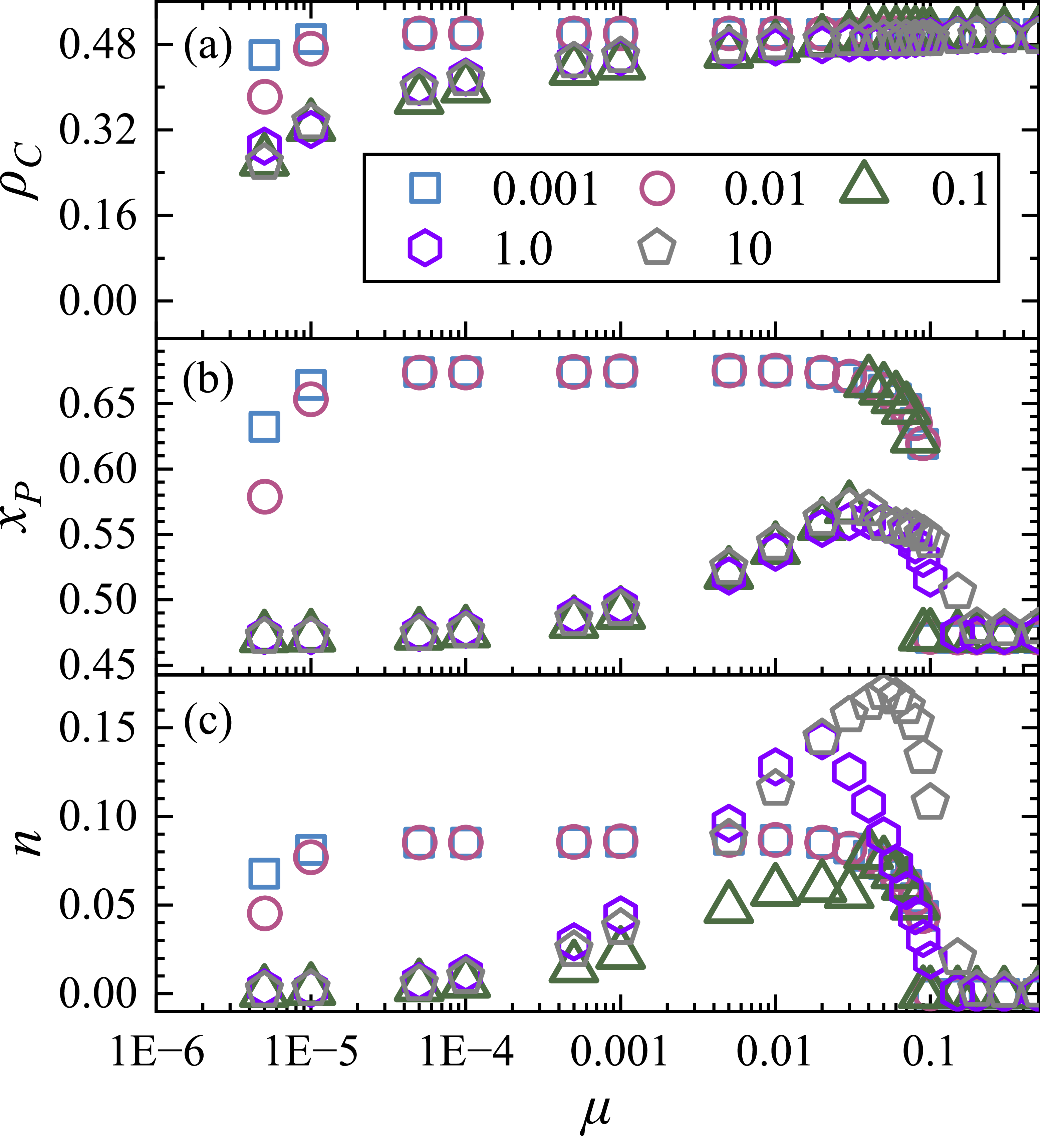}

  \caption{\label{fig_r03_3l}(a) cooperative fraction $\rho_C$, (b) population fraction $x_P$ and (c) environmental state $n$ as functions of mutation probability $\mu$ for $\epsilon=0.001$, $0.01$, $0.1$, $1.0$, and $10$. Each point is averaged over 30 repeats. } 
    
\end{figure}

\begin{figure}%[!htbp]

    \includegraphics[width=\linewidth]{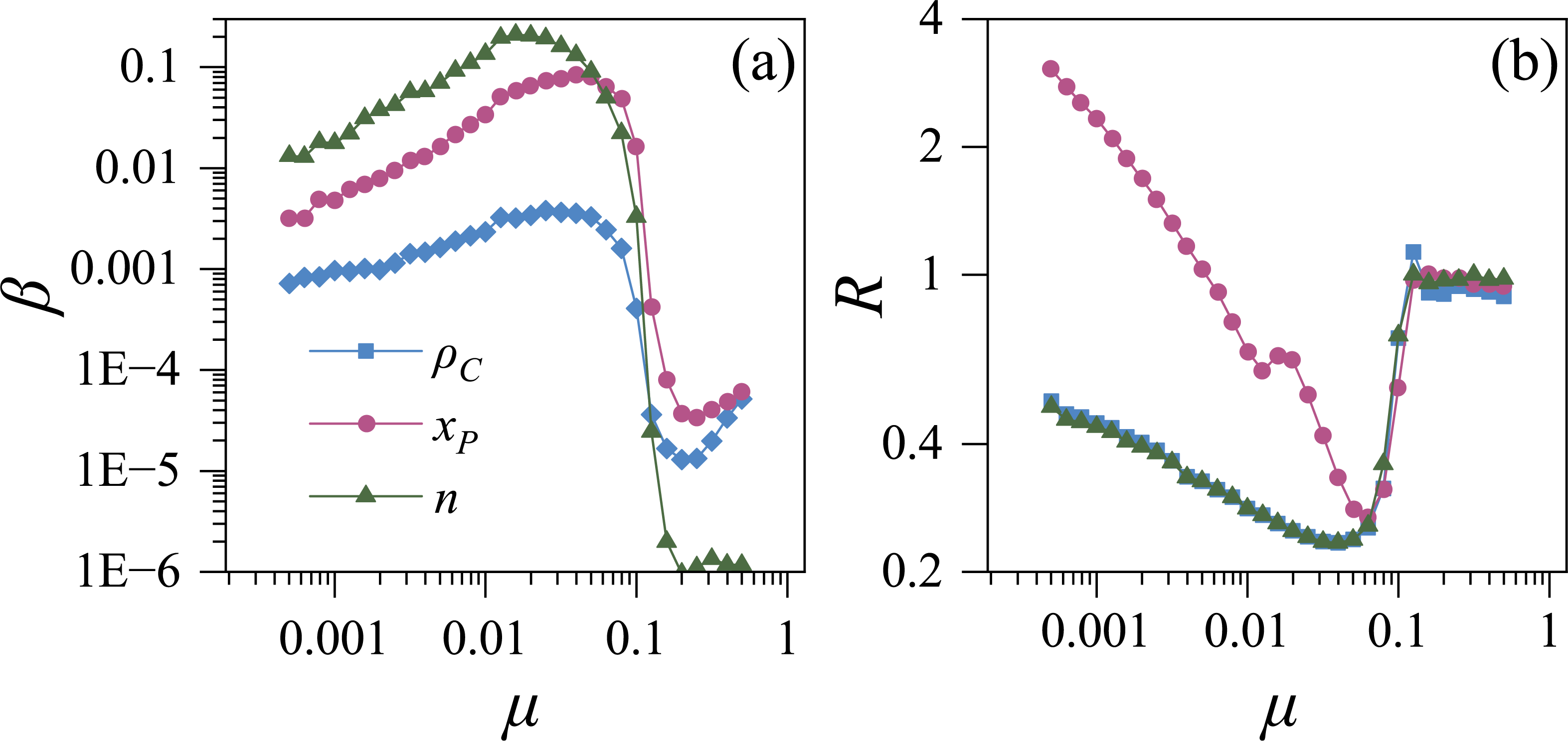}

  \caption{\label{fig_cr}(a) Degree of coherence $\beta$ and (b) coefficient of variation (CV) $R$ as functions of mutation probability $\mu$ for $\epsilon=1$. Three curves are calculated base on data of cooperative fraction $\rho_C$, population fraction $x_P$, and environmental state $n$, respectively. } 
    
\end{figure}

Our results indicate that environmentally and survival-friendly oscillations emerge in systems both with and without mutations, as illustrated in Fig.~\ref{fig_osc}. This figure displays two oscillatory processes, where the intervals between oscillations are unpredictable, and the oscillations persist even after extended evolutionary timescales.

We present seven snapshots of individual patterns during a typical oscillation process [Fig.~\ref{fig_osc}(a)--(g)]. Prior to the oscillation, cooperators or defectors are loosely clustered, with numerous empty nodes interspersed within the clusters [Fig.~\ref{fig_osc}(a)--(b)]. Thus, two types of regions are easily recognizable: cooperators with empty nodes, and defectors with empty nodes. In this phase, the death rate is maximized due to the poorest environmental conditions. The population remains at a low level, as individuals rapidly reproduce in the vicinity of empty nodes. We note here that our model allows for a habitable environment even at $n=0$, acknowledging that most real-world systems cannot be easily degraded to a state entirely incapable of supporting biological life.

As the population fraction rises, cooperators gradually cluster into tightly packed agglomerates because their death rate $\eta$ decreases as $n$ increases, while they mutually benefit from lower costs according to Eq.~\eqref{eq_payoff_matrix}. Meanwhile, the system generally evolves toward a survival-favorable environment state with the dominance of cooperators, accompanied by suitable environmental as indicated by Figs.~\ref{fig_osc}(c)--(e).

However, these cooperator clusters are susceptible to invasion by defectors, which leads to an increase in the defective population and subsequent environmental degradation. The population fraction decreases rapidly as the game's payoff diminishes and environmental damage intensifies, resulting in a higher death rate [Eq.~\eqref{eq_death_rate}]. Subsequently, the system returns to a state of environmental barrenness and low population, as shown in Figs.~\ref{fig_osc}(f)--(g). A gif illustrating the pattern evolution in oscillation processes is available in Ref.~\cite{cr_state_patterns}.

To further investigate the emergence of oscillations, we examined different mutation probabilities $\mu$ under several environmental update rates $\epsilon$. The results are presented in Fig.~\ref{fig_r03_3l}. For rapid environmental feedback ($\epsilon=1.0$ and $10$), population fraction $x_P$ and environmental state $n$ exhibit a single peak within $\mu \in (0.01, 0.1)$, forming bell-shaped curves. In contrast, the cooperative fraction displays a monotonically increasing trend as mutation probability $\mu$ increases.

The temporal evolution of all three variables exhibits similar oscillatory behaviors, as shown in Fig.~\ref{fig_osc}. Both $x_P$ and $n$ oscillate upward to peak values and then decline to persistently low levels. Conversely, $\rho_C$ oscillates upward to peak values, then drops below the values observed in non-oscillatory periods, and subsequently increases gradually back to its corresponding level. This phenomenon occurs due to the invasion of defectors following the oscillation peak, after which the cooperative population slowly recovers in the barren environment by mutually lowering the death rate. As a result, the average value of $\rho_C$ is minimally affected by the oscillations.

Since oscillations of $x_P$ and $n$ are always upward, their average values increase with the amplification of oscillatory dynamics, even though the peak values vary with oscillation frequency. Thus, the average $x_P$ and $n$ reach their highest peak values at intermediate mutation intensities $\mu$, as oscillations are optimally enhanced by noise in this regime. We refer to this phenomenon as survival coherence resonance. As shown in Fig.~\ref{fig_cr}, the peak in the degree of coherence at optimal noise intensity and the minimum in the CV confirm that this is a classical coherence resonance effect [Appendix \ref{app_cr}].

Previous studies have demonstrated that an optimal noise intensity maximizes the cooperative fraction in various evolutionary game scenarios, analogous to coherence resonance behavior \cite{traulsen2004stochastic,perc2006coherence,perc2006double,perc2006evolutionary,ren2007randomness,jia2017dependence,szabo2007evolutionary}. Our model reveals the upward oscillations of population fraction and environmental state as the environment evolves toward a suitable state, driven by the clustering and beneficial behavior of cooperators. Upon noise intensity optimization, high-frequency oscillations yield the highest peak values for the average $x_P$ and $n$. %之后加现实系统脉冲式快速波动的实例

\subsection{\label{sec_rd_des}Dynamic equilibrium state}

\begin{figure}%[!htbp]

    \includegraphics[width=\linewidth]{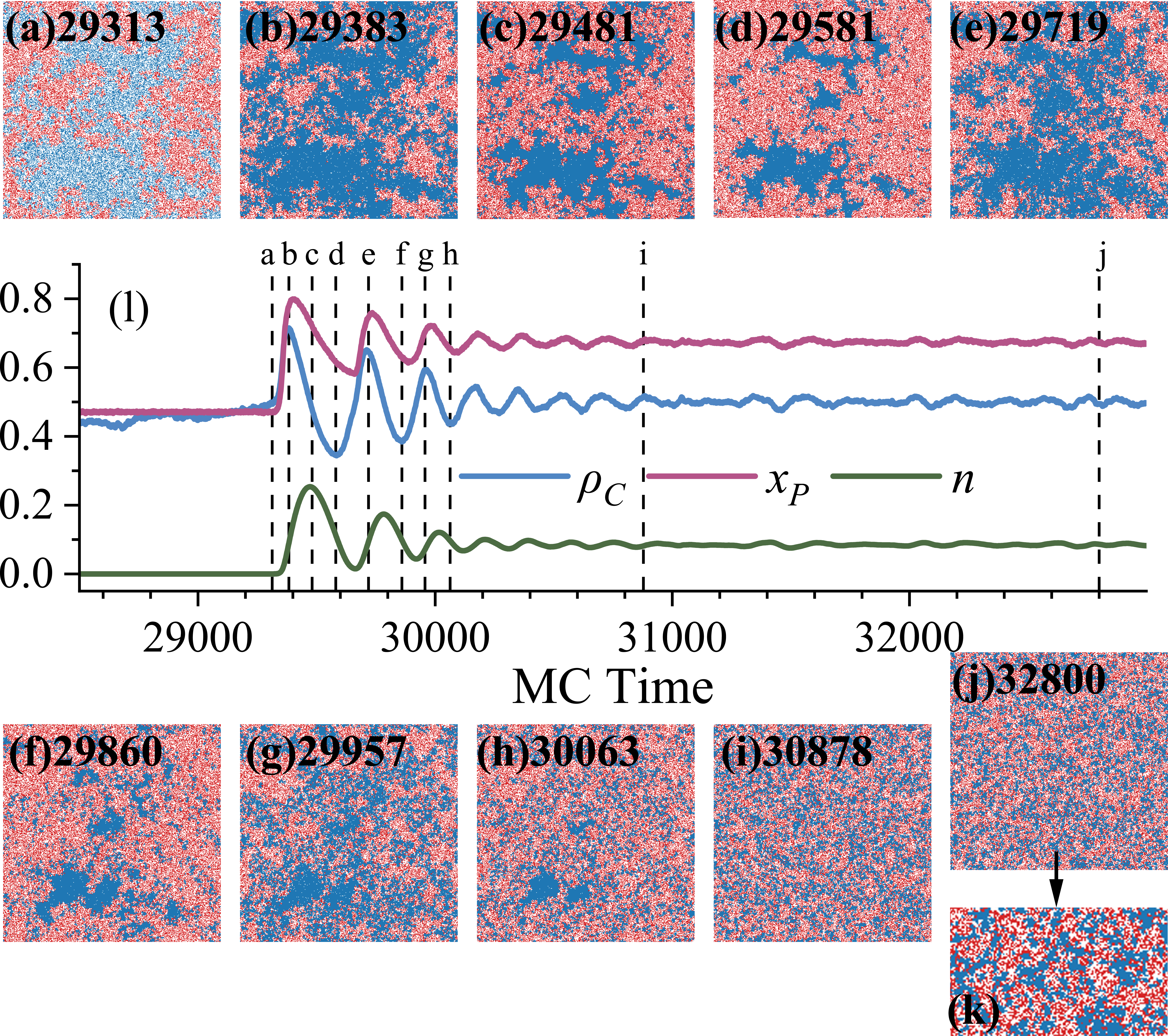}

  \caption{\label{fig_ds}Dynamic equilibrium state for $\epsilon=0.01$ and $\mu=0.00005$. (a)--(j) Spatial patterns of cooperators (blue points) and defectors (red points) at different times. (k)Localized enlargement of Panel (j). (l) Temporal evolution of the cooperative fraction $\rho_C$, population fraction $x_P$, and environmental state $n$. Dashed lines indicate the time points corresponding to panels (a)--(j). } 
    
\end{figure}

\begin{figure}%[!htbp]

    \includegraphics[width=\linewidth]{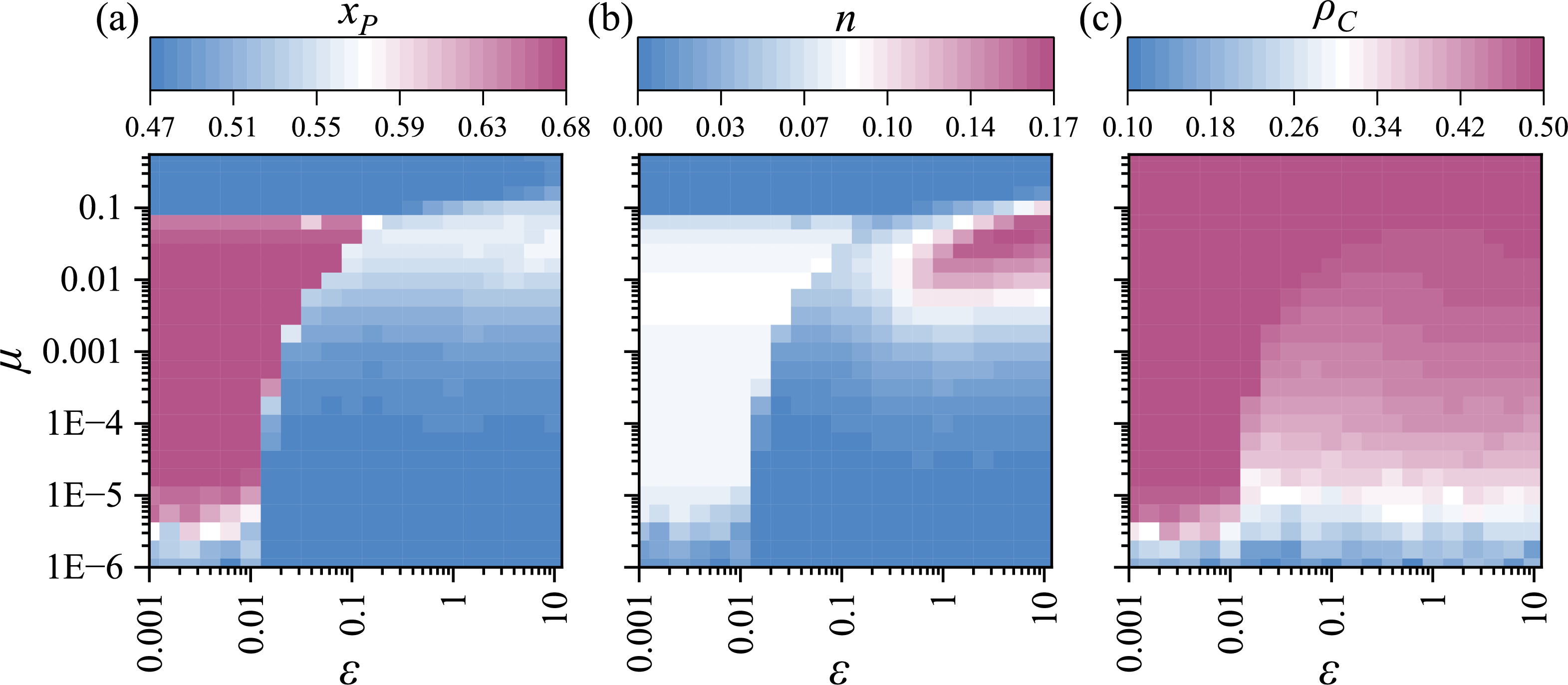}

  \caption{\label{fig_hm_p03}Heatmap of (a) population fraction $x_P$, (b) environmental state $n$ and (c) cooperative fraction $\rho_C$ in parameter space of $\epsilon$ and $\mu$. Each point in the heatmap is averaged over 30 repeats. } 
    
\end{figure}

As shown in Fig.~\ref{fig_r03_3l}, cases with lower environmental update rates ($\epsilon=0.001$, $0.01$, and $0.1$) exhibit higher peak values of $x_P$ across a wider range of mutation probabilities than the survival coherence resonance cases ($\epsilon=1.0$ and $10$). This occurs because the system enters a dynamic equilibrium state, characterized by self-organized tight clusters that maintain stable levels of population fraction and environmental quality. 

As shown in Fig.~\ref{fig_ds}, the system begins to enter the equilibrium state following the same upward oscillatory process as observed in the coherence resonance regime. After $\rho_C$, $x_P$, and $n$ reach their highest peaks, a damped oscillation process unfolds over time, and the system gradually stabilizes. However, it is important to note that reaching the dynamic equilibrium state is not guaranteed immediately after the system attains a peak. Under identical parameter conditions, the system stabilizes only after an unpredictable number of oscillations (i.e., after an indeterminate period of transient oscillatory dynamics).

The evolution of patterns intuitively reveals the mechanisms underlying the dynamic stabilization process \cite{equilibrium_state_patterns}. The system initially resides in a barren environment and low-population state, with both cooperators and defectors loosely clustered [Fig.~\ref{fig_ds}(a)]. The system first transitions into an upward oscillatory state with tightly clustered cooperators, as shown in Fig.~\ref{fig_ds}(b). As the environmental state reaches its peak, cooperators gradually lose their majority, and the population fraction decreases as the value of their games diminishes [Fig.~\ref{fig_ds}(c)].

Nevertheless, the cooperative fraction reaches a minimum while the environment remains favorable for the survival of cooperator clusters due to the slow environmental update rate [Fig.~\ref{fig_ds}(d)]. Cooperators are then able to reclaim territory from defectors, as the latter gain little benefit from clusters composed of defectors or empty nodes. Subsequently, the cooperative fraction recovers to a lower peak, which is followed by reduced peaks in both population fraction and environmental state. The system then becomes dominated by defectors, and the environment deteriorates again. This cycling process continues, with each peak decreasing and each valley increasing, as illustrated in Fig.~\ref{fig_ds}(e)-(i). Throughout these transitions, tight clusters persist, although their size fluctuates with the system state. Overall, cooperator clusters gradually become smaller and more uniformly distributed. Notably, the clustered patterns persist even after the system reaches the dynamic equilibrium state [Fig.~\ref{fig_ds}(i)-(k)].

A low environmental update rate naturally represents systems with high environmental carrying capacity, which cannot be easily depleted or enhanced. As shown in Fig.~\ref{fig_hm_p03}(a), the dark red region indicating higher population fractions is attributable to the dynamic equilibrium state with stable clusters. Each point in the heatmap is averaged over 30 repetitions. The bottom and top edges of the red region, featuring light red or white points, correspond to regions where survival coherence resonance and dynamic equilibrium state phenomena coexist. Correspondingly, the environmental state maintains low but positive values in this parameter region, as shown in Fig.~\ref{fig_hm_p03}(b), while the cooperative fraction also remains elevated [Fig.~\ref{fig_hm_p03}(c)].

The noise range for the dynamic equilibrium state narrows as the environmental update rate $\epsilon$ increases. This further suggests that the dynamic equilibrium state emerges when optimal noise induces cooperator clusters to withstand the system's fluctuating evolution toward equilibrium, forming smaller clusters without reaching the TOC state. In particular, cooperators can regain majority status as the population decreases alongside the environmental state. In systems with lower environmental capacity (higher $\epsilon$), the TOC state is reached again after each upward survival oscillation.

In summary, optimal noise induces the most favorable clustering behavior of cooperators, contributing to both survival coherence resonance and dynamic equilibrium state phenomena. The resonance phenomenon emerges for cost-to-benefit ratios $r$ near 0.3, which are slightly favorable for defectors. As the dilemma weakens or the environmental capacity is sufficiently high (i.e., lower $\epsilon$), the system attains a dynamic equilibrium state characterized by the presence of cooperative clusters more easily. See Appendix \ref{app_pd} for a detailed discussion of payoff structures.

\section{Conclusion}

In this work, we introduce dynamic environmental feedback into a three-state prisoner's dilemma game model on square lattices. The death rate is governed by individual payoffs determined by the static matrix, while the environmental state is regulated through strategic feedback. As the cost-to-benefit ratio $r$ increases, the dilemma becomes more pronounced, leading to a TOC state characterized by a barren environment and low population. Spatial structures such as the square lattices employed here enhance cooperation through the formation of mutually beneficial clusters of cooperators, resulting in a generally higher critical value of $r$ \cite{nowak1992evolutionary,szabo2007evolutionary}.

Environmental capacity determines the system's ability to withstand fluctuations. The system reaches and maintains an equilibrium state as cooperators cluster into mutually supportive agglomerates that both preserve environmental quality and resist invasion by defectors. A higher value of the environmental update rate $\epsilon$ naturally corresponds to lower environmental capacity. As $r$ increases, the system requires a lower $\epsilon$ to sustain an equilibrium state characterized by a high population fraction and a survival-favorable environment, or otherwise transitions to a TOC state with a barren environment and low population.

Upward oscillations of population arise as cooperators stochastically evolve into tightly knit mutualist clusters within the TOC state and barren environment. When cooperators dominate the system, the environment is restored to a level favorable for survival. However, this allows defectors to invade, driving the environment back to the TOC state. This process repeats irregularly over time and is most responsive to a specific optimal noise intensity. We refer to this phenomenon as survival coherence resonance, as both the average population fraction and the environmental state exhibit optimal peak responses. Due to these mechanisms, this type of resonance emerges at the boundary between cooperative and defective payoff structures, as well as at the transition between the abundant equilibrium state and the TOC state.

\appendix

\section{\label{app_pd}Influence of dilemma strength}

\begin{figure}%[!htbp]

    \includegraphics[width=\linewidth]{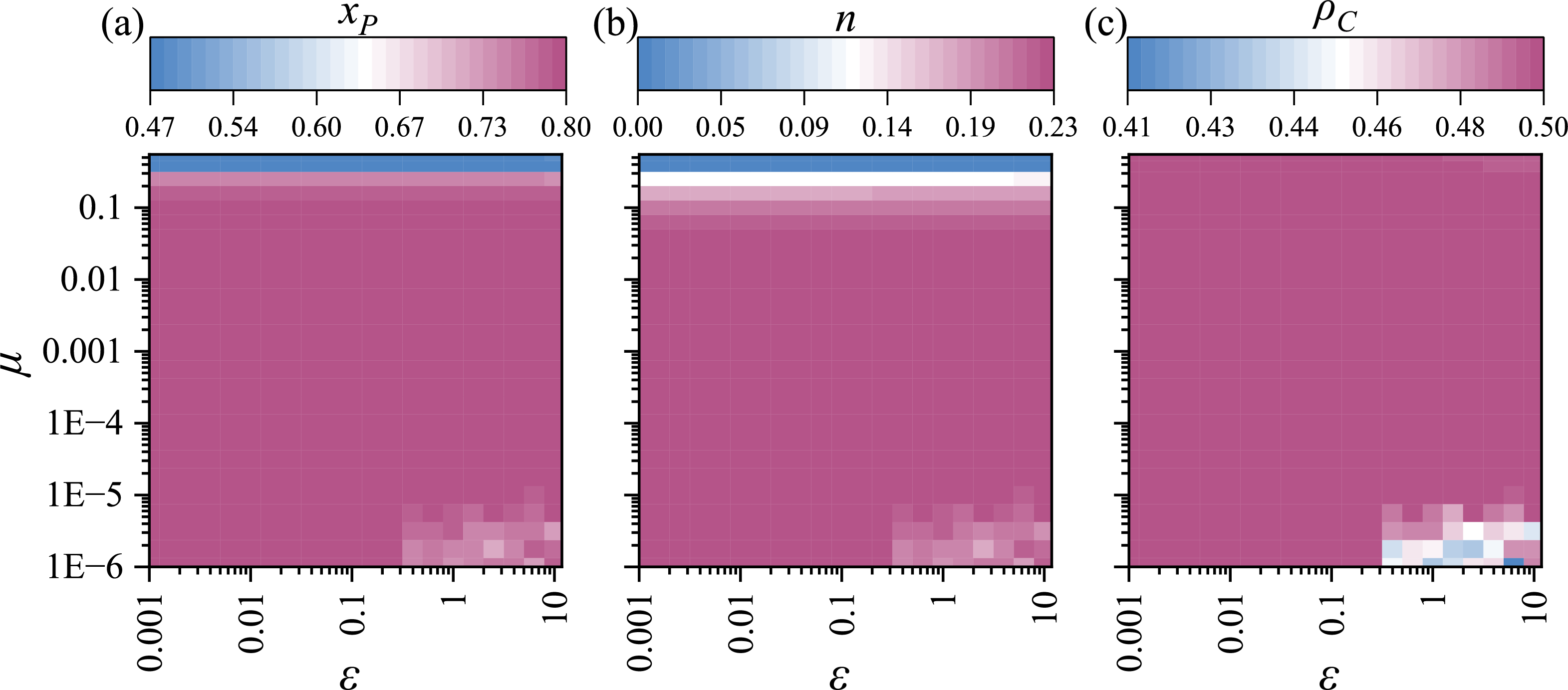}

  \caption{\label{fig_hm_p02}Heatmap of (a) population fraction $x_P$, (b) environmental state $n$ and (c) cooperative fraction $\rho_C$ in parameter space of $\epsilon$ and $\mu$ for $r=0.2$. Each point in the heatmap is averaged over 30 repeats. } 
    
\end{figure}

\begin{figure}%[!htbp]

    \includegraphics[width=\linewidth]{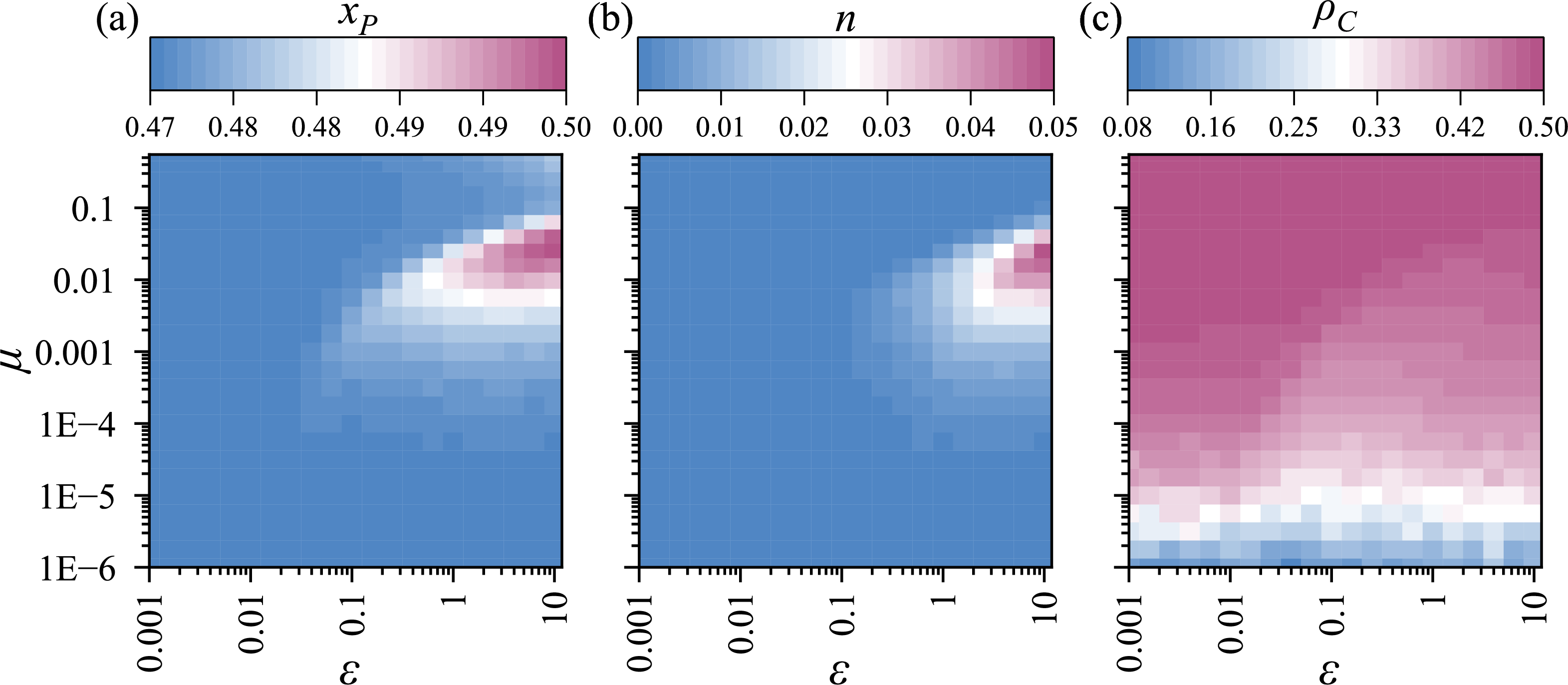}

  \caption{\label{fig_hm_p04}Heatmap of (a) population fraction $x_P$, (b) environmental state $n$ and (c) cooperative fraction $\rho_C$ in parameter space of $\epsilon$ and $\mu$ for $r=0.4$. Each point in the heatmap is averaged over 30 repeats. } 
    
\end{figure}

The cost-to-benefit ratio $r$ determines the strength of the prisoner's dilemma. The point $r=0$ marks the boundary between the harmony game ($r<0$) and the prisoner's dilemma game ($r>0$). It has been shown that square lattices enhance cooperation in the prisoner's dilemma \cite{nowak1992evolutionary,szabo2007evolutionary}. Consequently, as $r$ increases, the game becomes less favorable to cooperators until a positive critical value is reached, at which defectors dominate the system and a TOC state emerges under environmental feedback.

As shown in Fig.~\ref{fig_hm_p02}, the system maintains a high population fraction and a survival-favorable environment when mutation does not dominate the evolution of strategies. In this case, $r=0.2$ represents a weak dilemma, allowing cooperators to cluster easily and resist defection while preserving environmental quality. In general, values of $r$ lower than $r=0.3$ used in the main text lead to larger regions of equilibrium state with higher population fractions in favorable environments.

As shown in Fig.~\ref{fig_hm_p04}, $r=0.4$ results in no equilibrium state within the parameter space considered. Both the population fraction and environmental state are lower than in the case of $r=0.3$ presented in the main text. Further increases in $r$ (i.e., strengthening the dilemma) cause the coherence resonance phenomenon to gradually disappear. Thus, the upward oscillation of cooperator clustering that induces coherence resonance only occurs for $r$ values where the dilemma slightly favors defectors.

\section{\label{app_cr}Measurement of coherence resonance}

To investigate the coherence resonance phenomenon \cite{pikovsky1997coherence,lindner2004effects}, We calculate the degree of coherence $\beta$ and coefficient of coherence (CV) $R$ for each of the three variables $\rho_C$, $x_P$ and $n$ (where we refer to the the evolution of any of these variables over time as process $x(t)$, for convenience). The degree of coherence is defined as
\begin{equation}\label{eq_dc}
  \beta = \frac{S(\omega_{\mathrm{max}})}{\Delta \omega / \omega_{\mathrm{max}}},
\end{equation}
with 
\begin{equation}
  \begin{aligned}
    &\Delta \omega = \omega_2 - \omega_1, \\
    &S(\omega_1)=S(\omega_2)=S(\omega_{\mathrm{max}})/2, \\
    &\omega_1 < \omega_{\mathrm{max}} < \omega_2,
  \end{aligned}
\end{equation}
where $S(\omega)$ is the power spectrum of process $x(t)$, $S(\omega_{\mathrm{max}})$ is the maximum value, and $S(\omega_1)$ [$S(\omega_2)$] is the half-height position in the left (right) side. The CV is defined as 
\begin{equation}
  R=\frac{\sqrt{\langle \Delta T^2 \rangle}}{\langle T \rangle},
\end{equation}
where $\langle T \rangle$ is the mean interspike interval and $\langle \Delta T^2 \rangle$ is its variance. As previous studies have indicated, coherence resonance refers to the optimal system response at a specific noise level, resulting in the highest degree of coherence and the lowest value of CV. These two independent measurements together serve as reliable evidence for coherence resonance in this work (Fig.~\ref{fig_cr}).

% If you have acknowledgments, this puts in the proper section head.
\begin{acknowledgments}
    The authors warmly thank C. Liu at Hebei Normal University for helpful discussions. This work was supported by the National Natural Science Foundation of China (Grants Nos.~12575039, 11475074, 12247101 and 12375032), the Fundamental Research Funds for the Central Universities (Grant Nos.~lzujbky-2025-it50, lzujbky-2024-11, lzujbky-2023-ey02 and lzujbky-2024-jdzx06), the Natural Science Foundation of Gansu Province (No.~22JR5RA389), and the 111 Center under Grant No.~B20063. This work was partly supported by Longyuan-Youth-Talent Project of Gansu Province. 
\end{acknowledgments}

% Create the reference section using BibTeX:
%apsrev4-2.bst 2019-01-14 (MD) hand-edited version of apsrev4-1.bst
%Control: key (0)
%Control: author (72) initials jnrlst
%Control: editor formatted (1) identically to author
%Control: production of article title (-1) disabled
%Control: page (0) single
%Control: year (1) truncated
%Control: production of eprint (0) enabled
%

\end{document}